\documentclass[fleqn,usenatbib]{mnras}

\usepackage[T1]{fontenc}
\usepackage{siunitx}
\usepackage{newtxtext,newtxmath}

\DeclareRobustCommand{\VAN}[3]{#2}
\let\VANthebibliography\thebibliography
\def\thebibliography{\DeclareRobustCommand{\VAN}[3]{##3}\VANthebibliography}

\usepackage{graphicx}	
\usepackage{amsmath}	
\usepackage{amssymb}	

\title{Morpho-Kinematic Modeling of the Point-Symmetric Cat's Eye, NGC 6543: Ring-like Remnants of a Precessing Jet}

\author[R. Clairmont et al.]{
Ryan Clairmont$^{1}$\thanks{E-mail: ryanclairmont53@gmail.com}
Wolfgang Steffen,$^{2,3}$
Nico Koning$^{4}$
\\
$^{1}$Department of Physics, Stanford University,
382 Via Pueblo Mall,
Stanford, CA 94305, USA \\
$^{2}$Instituto de Astronomía, Universidad Nacional Autónoma de México, km 107 Carr. Tijuana-Ensenada, 22860 Ensenada, B. C., Mexico\\
$^{3}$ilumbra, AstroPhysical MediaStudio, Kaiserslautern, Germany\\
$^{4}$Department of Physics and Astronomy, University of Calgary, 2500 University Drive NW, Calgary, AB T2N 1N4, Canada
}

\date{Accepted XXX. Received YYY; in original form ZZZ}

\pubyear{2022}

\begin{document}
\label{firstpage}
\pagerange{\pageref{firstpage}--\pageref{lastpage}}
\maketitle

\begin{abstract}
 The planetary nebula known as the Cat’s Eye Nebula (NGC 6543) has a complex, point-symmetric morphology that cannot be fully explained by the current theory of planetary nebula formation, the Interacting Stellar Winds Model. In order to reveal the three dimensional (3D) structure of the Cat's Eye Nebula, we created a detailed 3D morpho-kinematic model of this nebula using a [NII] image from the Hubble Space Telescope and five different position-velocity diagrams using the SHAPE code. This modeling approach has revealed point-symmetric partial rings, which were likely formed by a precessing jet.
\end{abstract}




\section{Introduction}

Planetary nebulae (PNe) are formed from low to intermediate mass stars. Once the asymptotic giant branch (AGB) phase of a star ends, the star ejects its outer shell into the interstellar medium. The radiation from the hot, unveiled core of the AGB star photoionizes the ambient gas, producing a planetary nebula (\citealt{roxburgh1967origin}). The hot central star ejects a fast stellar wind that collides with a higher density environment that has been created by a slow dense stellar wind during the AGB phase. The fast wind shocks the dense gas, creating high thermal pressure that pushes a compressed shell outwards. The interaction of multiple wind ejection events that forms the structure of PNe is known as the interacting stellar winds model (ISW) (\citealt{kwok1978origin}). \par

NGC~6543, also known as the Cat’s Eye Nebula, is a spectacular planetary nebula renowned for its highly complex, point-symmetric morphology. It is bipolar, which according to the generalized interacting stellar winds model (GISW), is explained by an equatorial density enhancement in the gaseous medium around the nebula that hampers the expansion of the nebula’s outer shell at its equator (\citealt{balick2002shapes}).  \par

NGC~6543 contains a WR+Of central star which has a temperature between 40,000 and 70,000~K (\citealt{balick2004ngc}). A distance of 1.3 kpc was determined from GAIA parallax data (\citealt{chornay2021one}). Fast stellar winds from the central star have been observed by the International Ultraviolet Explorer (\citealt{perinotto1989fast}), and the Chandra X-ray Observatory observed soft X-rays emanating from the inner shell (\citealt{chu2001chandra}). \par

Its physical structure makes NGC~6543 perhaps one of the most complex and unusual planetary nebulae (\citealt{balick2002shapes}; \citealt{balick2004ngc1}). NGC~6543 contains two visible shells as seen in the image taken by the Hubble Space Telescope (HST) (Figure \ref{fig:slitpositions}). The inner shell is an ellipsoid, similar to those observed in NGC~7009 and NGC~2392 (\citealt{steffen20093d}; \citealt{garcia2012detailed}). The outer shell is bipolar, with two nearly spherical lobes connected at the center of the nebula. A pair of jets, ansae, and caps also appear around the nebula, all strikingly point-symmetric (\citealt{balick2004ngc1}).\par

In this study we do not include the halo (\citealt{Corradi2003}) which has a radius that is about ten times larger than what we refer to as the outer shell.\par 

\cite{miranda1992long} used spatio-kinematic modeling to determine that NGC~6543 contains an outer and inner ellipsoid and point-symmetric smaller features. They propose precessing collimated ejections as an explanation for the point-symmetry. We seek to expand upon their model by using the high resolution [NII] HST image of NGC~6543 as well as utilizing detailed 3D modeling. \par

Point-symmetry is not immediately explained by the GISW model unless additional modifications are made such as precessing outflows (\citealt{velazquez2012multipolar}) or warped disks (\citealt{rijkhorst2005}). A difficulty with the precessing collimated outflow scenario for point-symmetry is that it is not clear how the timescales on the order of years involved in the precession is transferred to the large scale evolution on a timescale of thousands of years and size scales of a parsec. Constant velocity ejections over a short time will merge on timescales of more than a hundred times this time scale. Conservation of a nebula's point-symmetry requires a precession timescale that is of the same order as the kinematic age of the largest parts of the structure. The latter is a problem, since the binary timescales that can influence the outflow direction are on the order of years. However, if the symmetric structures only comprise a distance range that is small compared to the size of the overall nebula, the timescale is reduced by the ratio between the distance range of the structure and the overall size of the nebula. \par

Some studies have suggested that NGC~6543 may have a binary central star (CS) due to its complex morphology and the variability of its spectral lines (\citealt{balick1987wind}; \citealt{miranda1992long}, \citealt{hyung2000optical}), although \cite{bell1994search} attributes the spectral line variability to variations in a single star. \par
     
The purpose of this study is to determine and visualize the 3D structure of the Cat’s Eye using morpho-kinematic modeling based on HST images and public spatially resolved kinematic data from ground-based longslit spectroscopy. \par

The paper is structured as follows. In section \ref{section:Observations} we describe the observations used to create and test the model. In section \ref{section:Kinematics}, we describe the relevant features seen in the PV data.  Section \ref{section:modeling} describes our modeling process. The resulting morpho-kinematic model and our findings are explained in section \ref{section:results} and the implications are further discussed in section \ref{section:discussion_conclusions}. \par

\section{Observations}
\label{section:Observations}
\begin{figure}
\includegraphics[width=\columnwidth]{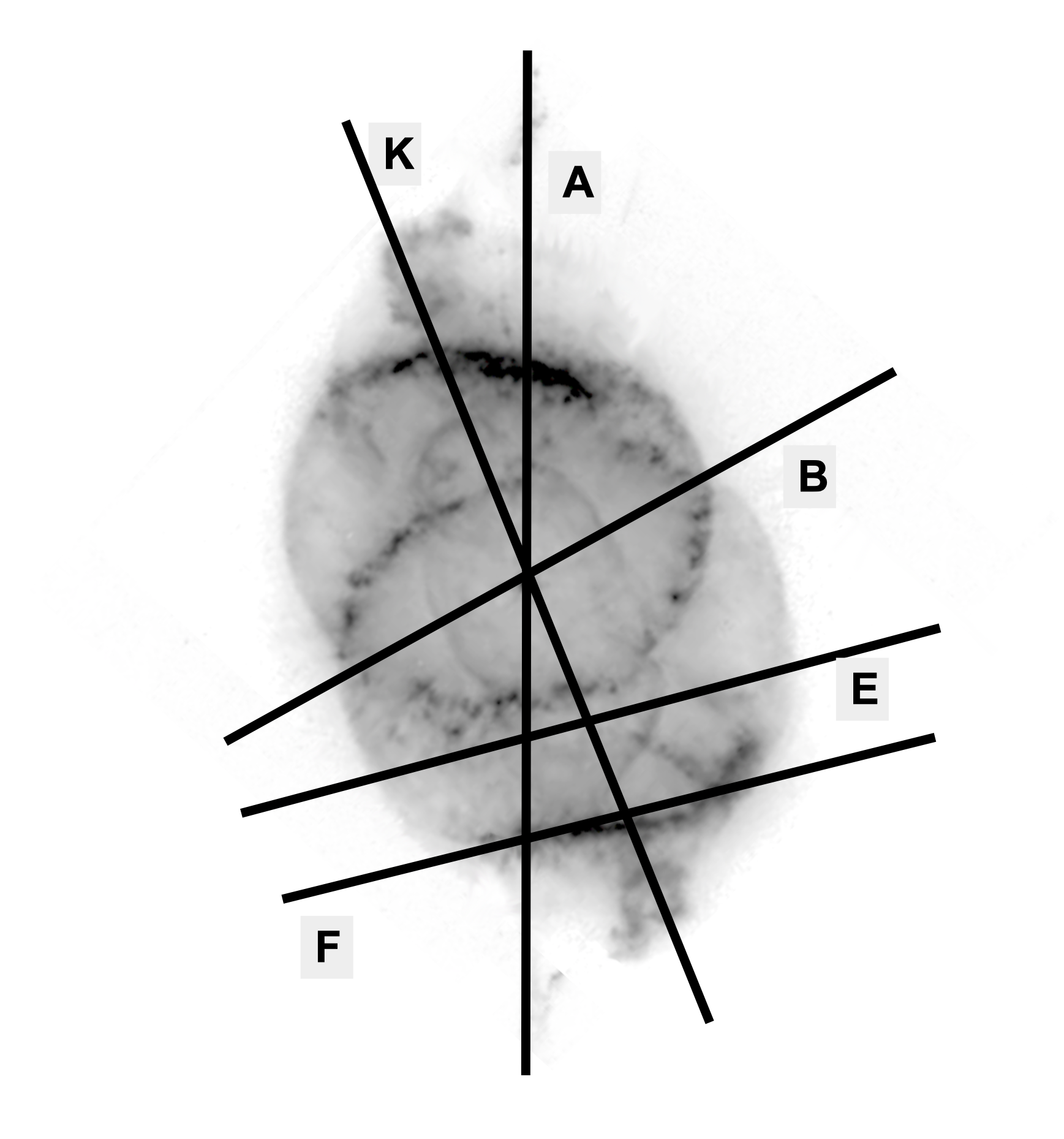}
\caption{Position of each slit shown on the HST image of NGC 6543. Each letter on this figure corresponds to the matching PV diagrams in Figure \ref{fig:rendered_pvs}.}
\label{fig:slitpositions}
\end{figure}

Five different position-velocity diagrams sampling the Cat's Eye were taken from the San Pedro Martir Kinematic Catalogue (\citealt{lopez2012san}), available at http://kincatpn.astrosen.unam.mx. The data were acquired using the Manchester Echelle Spectrometer (MES-SPM, \citealt{meaburn2003manchester}) on the 2.1 m, f/7.5 telescope at the San Pedro Martir National Observatory in Mexico. MES-SPM was equipped with a SITE-3 CCD detector with 1024 x 1024 pixels, resulting in 0.312 arcsec pixel$^{-1}$. For positions A, B, and K, a 150 $\mu$m slit was used, giving a velocity resolution of 11.9 km s$^{-1}$. For positions E and F, a 70 $\mu$m slit was used, resulting in a velocity resolution of 9.2 km s$^{-1}$. Using SAOImageDS9 (\citealt{joye2003new}), the spectral line for [NII] was isolated and put on a log scale, which made the fainter features visible in the PV diagrams. The PV diagrams' slit locations are shown in Figure \ref{fig:slitpositions}. In addition to the PV diagrams, an [NII] image taken by the WFPC2 camera on the HST was also used to construct the morpho-kinematic model. Due to the nitrogen higher atomic weight compared to hydrogen, the [NII] $\lambda$6584 has higher Doppler-velocity resolution. It also shows the best contrast for nebular features and better represents the knotty low-ionization structures distributed throughout the nebula (Balick 2004). The [NII] image was obtained from \cite{balick2004ngc1} and their data reduction process is described therein.\par

\begin{figure}
\includegraphics[width=\columnwidth]{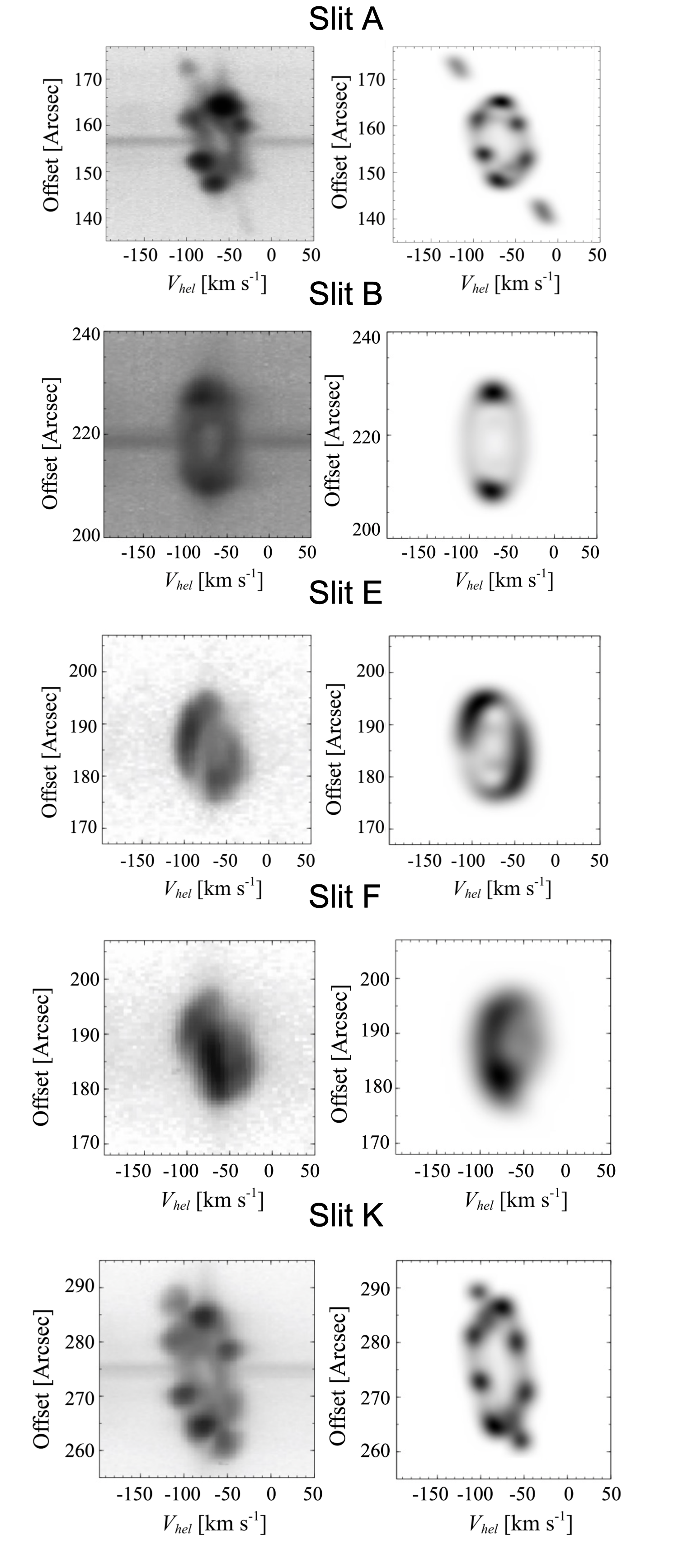}
\caption{Array of five Position-Velocity (PV) diagrams of NGC 6543. Synthetic PV diagrams generated from the morpho-kinematic model are on the right; observed PV diagrams are on the left. Observed PV diagrams were taken from the San Pedro Martir Kinematic Catalogue (\citealt{lopez2012san}). The comparison between the synthetic and observed maps allow the accuracy of the morpho-kinematic model to be tested.}
\label{fig:rendered_pvs}
\end{figure}

\section{Kinematics}
\label{section:Kinematics}

The [NII] PV diagrams are shown in Figure \ref{fig:rendered_pvs}. They include the observed PV diagram on the left, and the synthetic PV diagram on the right, generated from the SHAPE morpho-kinematic model. The heliocentric velocity  and the offset range varies for each diagram, but was chosen to include all of the visible features in the best detail, including the nebula's bipolar jets. We find that the nebula's heliocentric system velocity, $V_{sys} \simeq 65$ km s$^{-1}$ based on diagrams B and K, which both contain slits that go through the CS. \par

\subsection{Outer Shell}
The outer shell is visible in diagrams A, B, and K as an ellipse with a velocity of $V_{exp} \simeq 25$ km s$^{-1}$ at its waist. In slit K, which runs down the length of the nebula, the outer shell is slightly tilted, indicating that the long axis of the outer shell is tilted with respect to our line of sight. The north part of the ellipse is shifted towards blue, while the south part is shifted towards red. The HST image shows that the outer shell is pinched around the waist at the equatorial ring, but the ellipse in the PV diagrams does not show any variation in radial velocity at the waist. This could be because the resolution of the PV diagrams is not high enough to detect this small variation in radial velocity or because the equatorial ring is expanding at the same velocity as the rest of the outer shell. \par

\begin{figure}
\includegraphics[width=\columnwidth]{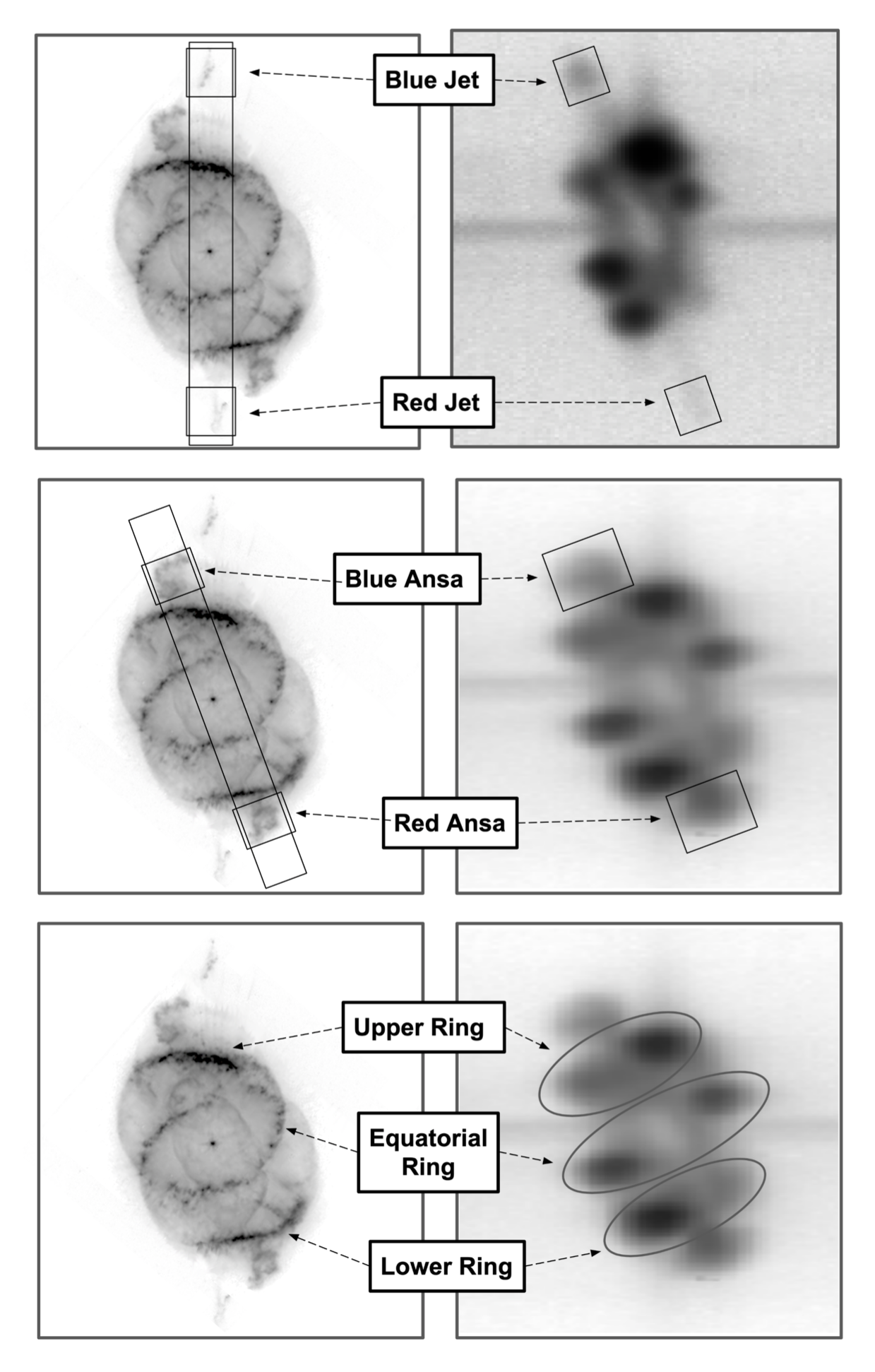}
\caption{Finding chart for identifying features on the PV diagrams (right) in the [NII] HST image (left). The corresponding features are labeled in each diagram.}
\label{fig:slit_guide}
\end{figure}

\subsection{Equatorial Ring}
The equatorial ring is visible in slits A, B, and K (see Figure \ref{fig:slit_guide}). It has approximately the same expansion velocity as the outer shell ($V_{exp} \simeq 25$ km s$^{-1}$). The equatorial ring is positioned at the center of the ellipse in the PV diagrams, which supports that the outer shell is bipolar. 

\subsection{Inner Shell}
The inner shell is not visible in any of the PV diagrams. This could be due to limited resolution. The inner and outer shell could blend together in the diagrams, especially if the inner shell and outer shell have similar expansion velocities. For the purposes of this study, we mainly relied on the HST image to construct the inner shell in the morpho-kinematic model. 

\subsection{Ansae}
The ansae are visible in slits A and K.  Taking into account that $V_{sys} \simeq 69$ km s$^{-1}$ , $V_{r} \simeq 25$ km s$^{-1}$ for the ansae. The ansae are strikingly point symmetric in the PV diagrams.  A line connecting the opposite ansae is parallel and right on top of the ellipse's long axis in slit K, which indicates that the ansae lie along the same axis as the outer shell, at its poles.

\subsection{Bipolar Jets}
The bipolar jets are only visible in slit A. They have a radial velocity of $V_{r} \simeq 35$ km s$^{-1}$. The jets can be traced from the outer shell to an extent of $\approx$19 arcseconds away from the CS. The tip of each jet is much brighter than the body of the jet in the PV diagrams and in the HST image, which suggests that the tip of the jet contains more shock-excited material than the body of the jet.

\subsection{Partial Rings}
The two partial rings are visible in slits A, E, F, and K. One ring is located north and the other south of the nebular waist equator. The rings are point-symmetric to each other and tilted at approximately the same angle, which also matches the tilt angle for the equatorial ring. Both rings are also equidistant from the nebular waist.

\section{Modeling in SHAPE}
\label{section:modeling}
\subsection{Morpho-Kinematic Model}
\label{section:morphokinematics}

A detailed morpho-kinematic model is derived in the astrophysical code SHAPE (\cite{steffen2006morpho}, \cite{steffen2013wind}, \cite{STEFFEN201787}) to clearly reveal and visualize the 3D structure of the Cat’s Eye Nebula. The morpho-kinematic model is constructed based on kinematic spectral data and high resolution images of the nebula. SHAPE has been used previously to model planetary nebulae similar to NGC~6543, including NGC~2392 (\citealt{garcia2012detailed}) and NGC~7009 (\citealt{steffen20093d}). \par

The workflow of SHAPE is as follows: Primitive mesh objects (cylinders, spheres, etc.) are set up and modified interactively to yield the complex structures seen in planetary nebulae. This part of the process is similar to modeling in conventional 3D animation packages. The volumes enclosed by the meshes serve as containers of the volumes occupied by the parts of the nebula. They are then assigned physical properties such as a density distribution, a velocity field and an emissivity. The emissivity is set to be proportional to the square of the density, since the emission stems from recombination. In the rest frame, the spectral line shape is set to be a Gaussian as a function of wavelength.
Synthetic images and position velocity (PV) diagrams are generated from the morpho-kinematic model and qualitatively compared to the observed counterpart. The user then modifies the structure and velocity of the morpho-kinematic model until a satisfactory synthetic representation of the observed images and PV diagrams is achieved. The comparison between observation and model is done by visual inspection when superimposing and blinking between them within SHAPE. A numerical criterion, such as least-squares, for finding the best fitting model is not done. \par
     
A homologous expansion law for individual shells or other features was implemented to create the model, which assumes that the velocity vector is proportional to the position vector. Most planetary nebulae approximately exhibit homologous expansion for individual shells, as evidenced by the similarity of their position-velocity diagrams to their physical structures (\citealt{steffen20093d}). For most of the nebula a single homologous expansion law was used. However, the inner ellipsoidal shell was not clearly detected in the PV\-diagrams and is likely to expand faster than the main shell features, since it expands into the low-density interior of the main bubble and is still filled with hot gas, as evidenced by the x-ray emission. Its structure was constrained with the imaging data and based on the assumption that it is aligned with the overall axis of the main nebula.  \par

\section{Results}
\label{section:results}
\subsection{3D Structure of the Cat's Eye}
\begin{figure}
\includegraphics[width=\columnwidth]{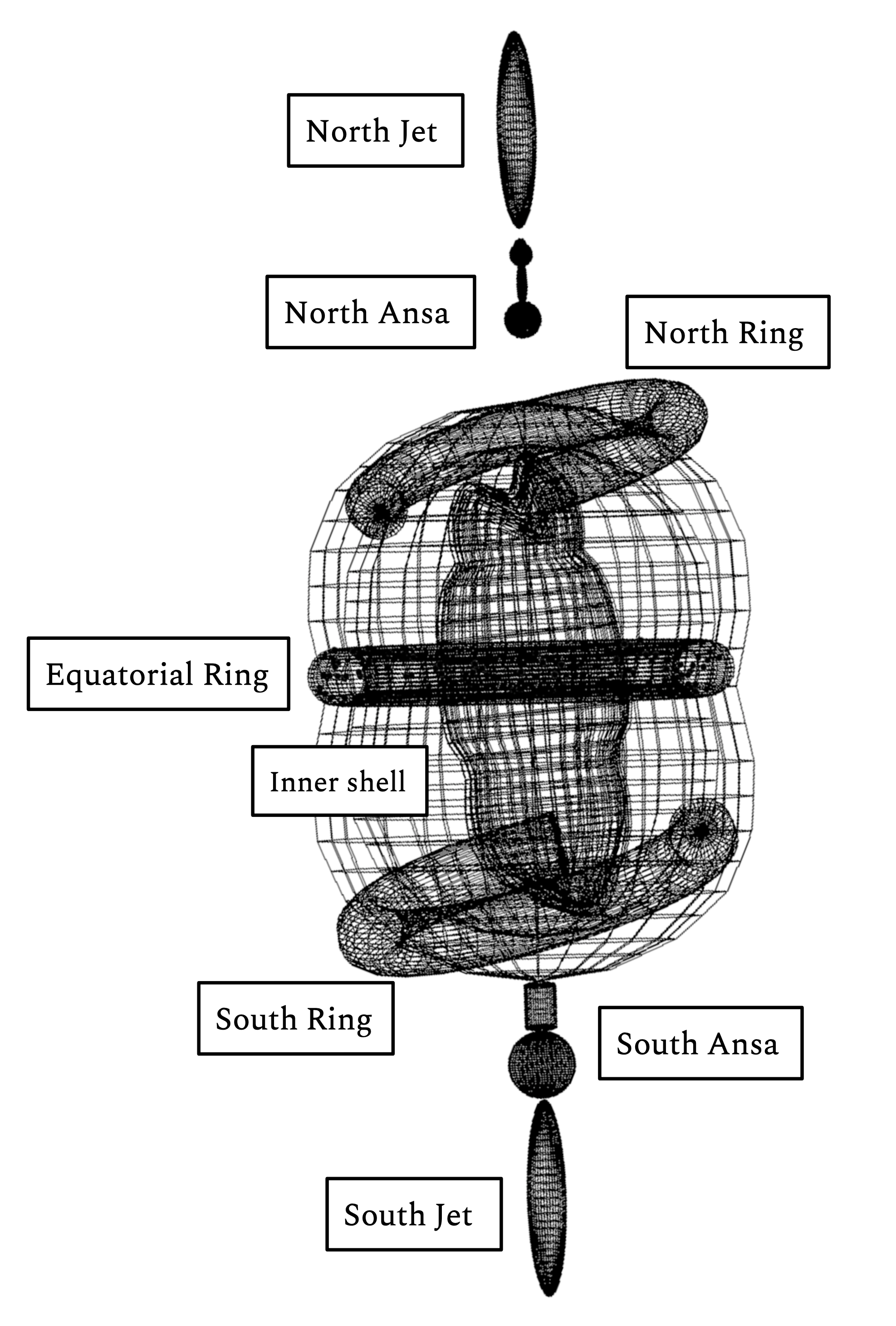}
\caption{SHAPE mesh model of the Cat's Eye as viewed from the side. The main components of the model are labeled.}
\label{fig:model_mesh}
\end{figure}

\begin{figure}
\includegraphics[width=\columnwidth]{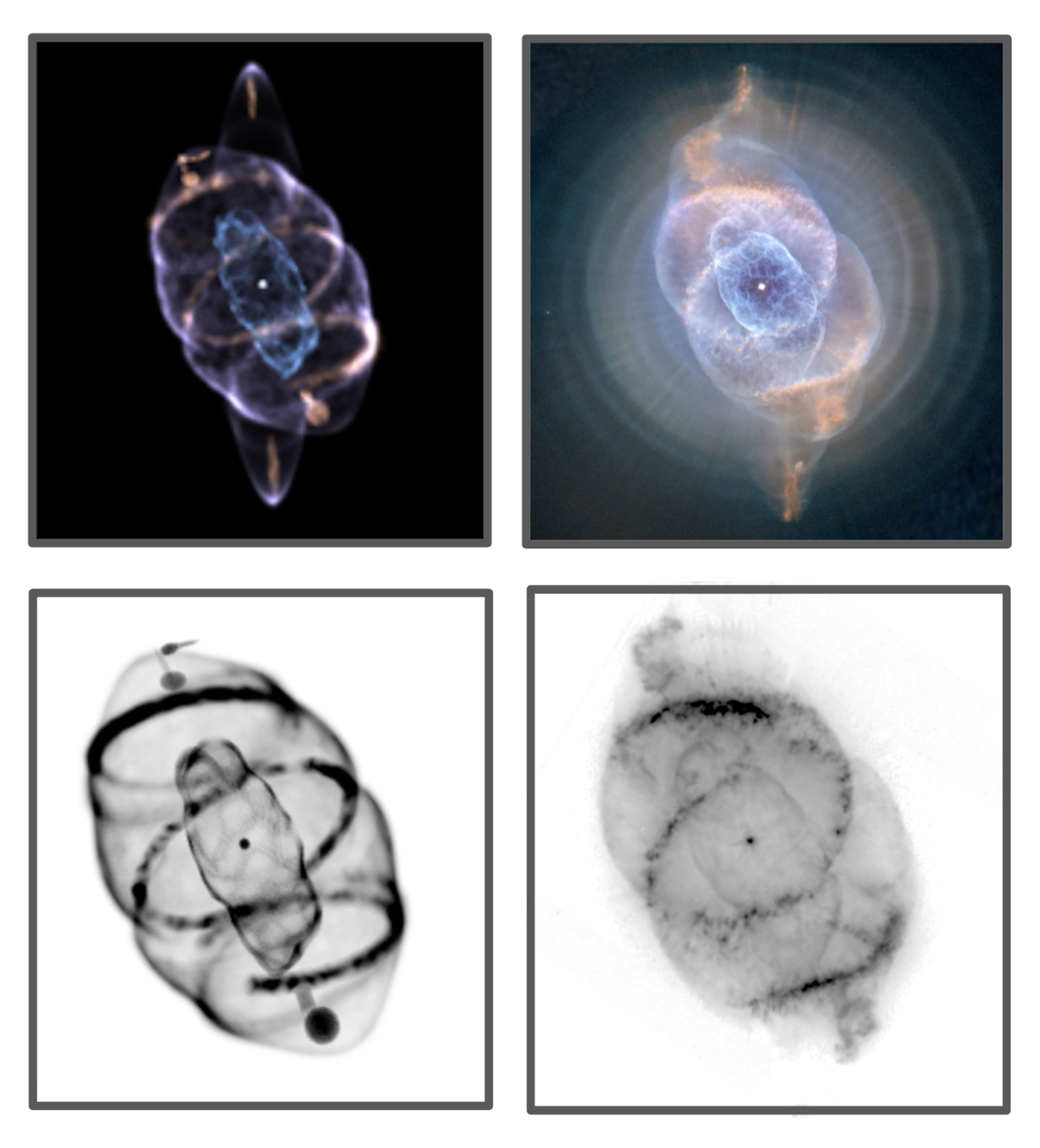}
\caption{Morpho-Kinematic Model in color (top left) to match HST color image (top right). [NII] Morpho-Kinematic Model (bottom left) to match HST [NII] image (bottom right).}
\label{fig:rendered_obs_model}
\end{figure}

The morpho-kinematic model reveals the 3D structure and, to some extent, the velocity distribution of most of the features of the nebula. The mesh rendering of the morpho-kinematic model is shown in Figure \ref{fig:model_mesh}, while Figure \ref{fig:rendered_obs_model} shows the morpho-kinematic model compared to the HST image. \par

The five PV diagrams along with the high resolution HST image allow the nebula's general structure to be reconstructed. The PV diagrams show that the outer shell is radially expanding and that the nebula's shells generally follow a homologous, radial expansion law since these features have the same relative positions in both the PV diagrams and the HST image. We used symmetric circular rings on the morpho-kinematic model to match the dark spots on the PV diagrams. In slits A and K, the clumps (upper, equatorial, and lower rings) show a slightly larger velocity dispersion than the clumps in the synthetic PV diagrams. This could be due to shock-driven acceleration of the rings at the interface between the rings and the expanding shell, which is not accounted for in the morpho-kinematic model.  \par

The inner shell cannot be separately observed in any of the PV diagrams, so only the HST image was used for reference when modeling the inner shell. In [NII], the inner shell is not bright and may be too thin to be readily visible in the PV diagrams with ground-based spatial resolution. It is also possible that the emission from the outer shell completely covers the inner shell, making it virtually invisible in the PV diagrams. The [NII] image from the HST barely shows the inner shell, and only by putting the image on a logarithmic scale is the inner shell readily visible (Fig 5). Alternatively, the inner shell may also have a very high shock-driven velocity (as evidenced by its X-ray emission), so the inner shell may actually appear larger than the outer shell in the PV diagrams and remain undetected. \par

The main finding of this paper is the point-symmetric partial rings around the outer shell of the Cat's Eye. \cite{miranda1992long} sketched a spatio-kinematic model to identify high density features in the PV diagrams without attempting to reproduce the observed PV diagrams. They identify the inner shell, outer shell, and equatorial ring, which we reproduce here. \cite{miranda1992long} attribute the partial rings on the PV diagrams as "polar caps" or "filamentary structures," but in this study we show that these features are actually point symmetric partial rings. \par

Additional PV diagrams in \cite{miranda1992long} show the radial velocities of the rings at multiple different position angles. The radial velocity variations along the ring are consistent with uniform expansion.  The part of the ring that faces towards Earth has a large negative radial velocity (-30 km s$^{-1}$ > $V_{r}$ > -36 km s$^{-1}$). As the ring loops around towards the side facing away from Earth, the radial velocity approaches zero, then becomes positive (30 km s$^{-1}$ < $V_{r}$ < 36 km s$^{-1}$). Additionally, the radial velocities of the top and bottom ring are point-symmetric: opposite points on each ring have approximately the same radial velocity, suggesting that they are both similar structures formed by the same mechanism. \par

We interpret both rings as only partial rings ($\ang{0} < \theta < \ang{300}$). The red-shifted half of the ring is only connected to the blue-shifted half of the ring at one point. This is demonstrated by the radial velocity variations in \cite{miranda1992long}, which show that when the red-shifted and the blue-shifted half of the same ring connect, the radial velocity approaches zero. On the opposite side, the blue and red-shifted halves' radial velocities do not approach zero. The radial velocity difference of both halves at the point where they should connect is 39 km s$^{-1}$ on the bottom ring and 35 km s$^{-1}$ on the top ring, which suggests that there is a spatial gap between the red and blue-shifted halves of the ring at that point (i.e., the ring is only partial). The spatial gap in both rings is also point-symmetric.

\section{Discussion and Conclusion}
\label{section:discussion_conclusions}

The 3D morpho-kinematic modeling approach used in this paper yields the following data for the Cat's Eye: 
\begin{itemize}
    \item  $V_{exp} \simeq 33.5$ km s$^{-1}$ at the poles of the outer shell.
    \item $V_{exp} \simeq 25$ km s$^{-1}$ at the equator of the outer shell.
    \item $V_{exp} \simeq 24 $ km s$^{-1}$ for the equatorial ring.
\end{itemize}

In accordance with the GISW model, the outer shell can be explained by the AGB superwind colliding with the ambient medium, creating the outer shell's sharp edge. This is similar to the outer shell in NGC 6826 (\citealt{plait1990evolution}). \par

The equatorial ring likely resulted from an equatorial density enhancement which pinched the nebula at its waist according to the GISW model (\citealt{balick2002shapes}). Using expansion parallax observations, \cite{balick2004ngc1} measured that the equatorial ring expands $70\%$ faster than the rest of the nebula. However, the PV diagrams show that the equatorial ring expands at approximately the same speed as the outer shell.

Expansion parallax observations only measure the expansion of the outermost shock front, while the PV diagrams show the expansion of the bulk mass of nebular material (\citealt{balick2004ngc1}). \cite{mellema2004expansion} calculated that the shock front travels 20-30\% faster than the post-shocked material behind it, leading expansion parallax measurements to overestimate the nebular expansion rate.  This alone does not account for the discrepancy, since the observations by \cite{balick2004ngc1} showed that the shock speed of the equatorial ring is 70\% faster than the shock speed of the rest of the nebula (i.e., both measure shock speed).  \par

Since the equatorial ring lies at the interface between two expanding bubbles (the bipolar lobes), \cite{balick2004ngc1} suggest that the equatorial ring has broken up into knots due to Rayleigh-Taylor instabilities after it encountered an equatorial density enhancement in the ISM. These knots are visible around the leading edge of the equatorial ring in the HST image. The instabilities at the edge of the equatorial ring allow outside material to flow through the ring as it expands. While the rest of the outer shell is decelerated by colliding with ambient material, the equatorial ring's shock front would continue to expand at the same speed, since it does not ram the ambient material. Because the permeable shock front is likely thin compared to the entire mass of the equatorial ring, the bulk material in the equatorial ring is still decelerated and expands with the rest of the outer shell, as seen in the PV diagrams. \par

Using the most recent distance measurement of 1.31 kpc by \cite{chornay2021one} and the homologous velocity law in our morpho-kinematic model, the kinematic age of the model was found to be 2,136 years. Since our model was based on a homologous velocity law, all the components have the same kinematic age, so we cannot suggest a particular evolutionary sequence solely based on the model. \par

The inner shell is a prolate ellipsoid nested inside the outer shell. Since we could not clearly distinguish the inner shell from the outer shell in the PV diagrams used in this study or in those acquired by \cite{miranda1992long}, we only used the HST image to model the inner shell assuming that it has an approximately cylindrical symmetry with minor local deviations. These local deviations from symmetry are evident on the poles of the inner shell. They may be caused by pockets of high density gas inside the outer shell that deform the inner shell as it expands. The ellipsoidal structure of the inner shell could be a result of inwards pressure exerted by the equatorial ring on the inner shell, constraining its equatorial expansion, as seen in the PV diagrams.    \par

The bipolar jets are modeled as ellipsoids positioned on an axis rotated $\ang{26}$ from the main axis of the nebula. It is the Cat's Eye's fastest component, with a deduced expansion velocity of $V_{exp} \simeq 72$ km s$^{-1}$. The bipolar jets are probably one of the oldest features of the nebula. \cite{miranda1992long} suggest that the bipolar jets are one of the newest features of the nebula and that they formed after the outer shell. However, it seems unlikely that the jets could have broken through the boundary of both the inner and outer shell without being significantly deformed. Additionally, supposing that the Cat's Eye contains a binary CS, the ejection of the hot, fast wind predicted by the ISW model would most likely shut off mass transfer in the binary, ending any collimated outflows.  \par

The ansae are modeled as a sphere for the lower ansa and an ellipsoid for the upper ansa. Both are located on the nebula's symmetry axis, right outside of the outer shell at its poles. Their expansion velocity is $V_{exp} \simeq 44$ km s$^{-1}$. There is no evidence in the PV diagrams or HST image that the ansae have a continuous stream of material originating from the CS, so they probably resulted from a singular ejection episode. \par

The partial rings on the nebula are modeled as partial tori. The northern ring has an expansion velocity of $V_{exp} \simeq 19.5$ km s$^{-1}$ with respect to the ring's center, while the southern ring has an expansion velocity of $V_{exp} \simeq 21.5$ km s$^{-1}$ with respect to the ring's center. Both rings have approximately the same expansion velocity with respect to the CS as the outer shell, since they are wrapped around the outer shell at its surface.  \par

\begin{figure}
\includegraphics[width=\columnwidth]{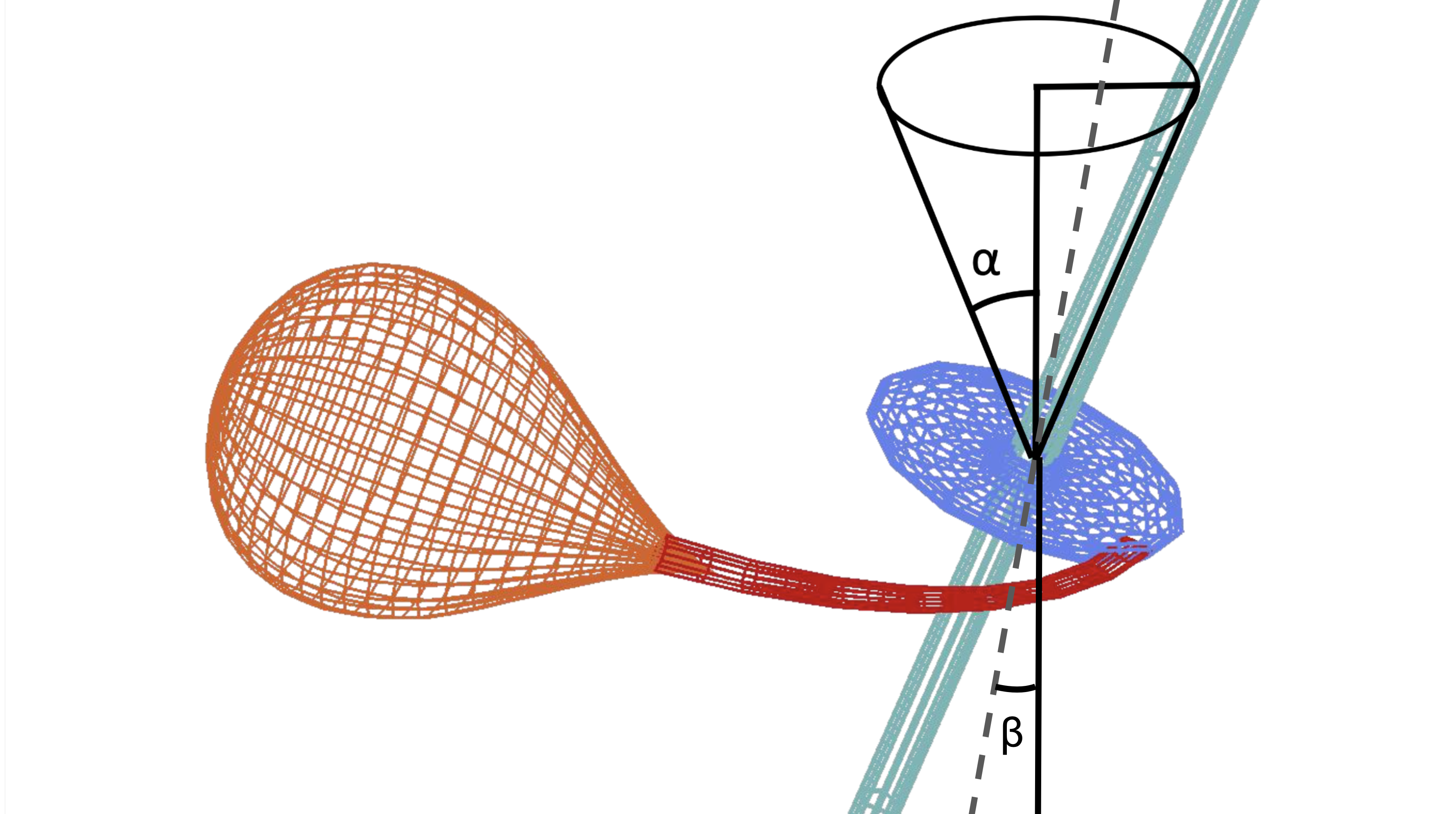}
\caption{The precession of an accretion disk around the secondary causes the jet to precess at an angle $\alpha$, where $\alpha \simeq \ang{38}$. The dashed line is the symmetry axis of the nebula. The rings are tilted away from the symmetry axis by $\beta \simeq \ang{17}$, which is the tilt angle of the precessing jet that formed the ring-like structures on the Cat's Eye.}
\label{fig:binary_diagram}
\end{figure}

These partial rings were likely formed by a precessing jet. To determine the precession angle, we imagined that each ring formed the base of a cone with the tip of the cone at the CS (see Figure \ref{fig:binary_diagram}). We measure a half-opening angle of $\alpha \simeq \ang{38}$ for both rings, which implies the precessing jet's half-opening angle is $\ang{38}$. Both rings are tilted an angle of $\beta \simeq \ang{17}$ with respect to the nebula's axis of symmetry, which suggests that the precessing jet was tilted at this angle with respect to the main symmetry axis of the central object when it formed the partial rings. \par

The precessing jet which likely created the partial rings never completed a full $\ang{360}$ rotation, resulting in the creation of only partial rings. The precessing jet scenario we envision at the level of the CS is shown in Figure \ref{fig:binary_diagram}. For a structure created by a precessing jet, one would expect there to be a noticeable pitch angle. However, from the model the pitch angle from the partial rings is small but not well determined, since only a fragment exists of the expected helix. It is also possible that short-term revolutions of the precessing jet converged on a larger timescale to form the partial ring seen in the Cat's Eye. The rings were likely formed before the outer shell. As the superwind was ejected from the binary CS, mass transfer ceased, causing the precessing jet to shut off. The superwind created the expanding boundary of the outer shell, which likely caught up with the partial rings. The partial rings can be interpreted as analogs to other ring-like filamentary structures formed by precessing jets in planetary nebulae, such as those seen in IC 4634 (\citealt{guerrero2008multiple}) or Fleming 1 (\citealt{boffin2012interacting}).

\subsection{Existence of a Binary Star In the Cat's Eye}
The 3D structure revealed by the morpho-kinematic model strongly indicates that a binary central star played a significant role in shaping the structure of the Cat’s Eye. Binaries have been often used to explain collimated outflows and point-symmetric, precessing jets (\citealt{jones2017binary}; \citealt{imai2002collimated}; \citealt{miszalski2009binary}) and have been discovered at the center of many planetary nebulae (\citealt{jones2017binary}; \citealt{brown2019post}).

Spectroscopic studies disagree on whether there is a close binary CS in the Cat’s Eye or simply a variable single star (\citealt{hyung2000optical}). However, the morpho-kinematic model presented in this study is a strong indicator of a close binary CS. It has been found through theoretical and numerical hydrodynamic studies that single stars do not have the necessary angular momentum to collimate winds and drive precessing jets or to produce large-scale deviations from spherical symmetry in PNe (\citealt{jones2017binary}; \citealt{imai2002collimated}; \citealt{miszalski2009binary}). The precessing jets, collimated winds, and large deviation from spherical symmetry shown in our models supports the idea that a close binary CS shaped this nebula’s unusual, complex structure.\par

\subsection{Proposed Evolution of the Cat's Eye}

Based on current theories of planetary nebulae evolution and the results of our morpho-kinematic model, we hypothesize that the evolution of the Cat’s Eye could be as follows: 

In a binary system containing an AGB star and a secondary star, the AGB star ejected an isotropic wind with a periodic change in the wind ejection rate, creating the spherical shells seen in the halo of the HST image (\citealt{balick2001ngc}). The binary’s gravitational pull on the spherical wind (\citealt{bermudez2020agb}) created a moderate equatorial density enhancement in the ambient medium (\citealt{balick2002shapes}). 

At some point, the AGB star overflowed its Roche lobe and began mass transfer through the inner Lagrangian point to the secondary. An accretion disk formed around the secondary as a result of the mass transfer. Magnetic fields generated by the accretion disk collimated wind from the secondary (\citealt{balick2020models}), causing two sets of dense clumps to be ejected at different times, forming the ansae. The difference in the clumps’ ejection angle is presumably due to the precession of the accretion disk, which also caused the subsequent jet to precess (\citealt{miranda1992long}, \citealt{sahai2005sculpting}). 

Then, the mode of mass-loss changed as the AGB star ejected a superwind, which became collimated due to the equatorial density enhancement (\citealt{balick2002shapes}), forming the nebula’s outer bipolar shell. The formation and ejection of the common envelope (CE) probably occurred after the ejection of the precessing jet, since the CE would likely block the jet ejection from the CS. 

The inner shell was formed by the shock-collision of a hot, fast wind predicted by the ISW model with material inside the outer shell. Alternatively, supposing that the binary CS formed a circumbinary excretion disk, the inner shell could have been formed by the interaction of a photoevaporated wind from the excretion disk with a fast wind from the CS (\citealt{garcia2018common}).\par

\subsection{Conclusions}
 A detailed 3-D morpho-kinematic model was constructed based on high-resolution imaging and spatially resolved kinematic observations of NGC~6543. Our model reveals partial rings that are likely remnants of a precessing jet. We also suggest a possible sequence of ejections events based on the hypothesis of a binary central star and its evolution that explain highly complex structures such as those of NGC~6543 and similar planetary nebulae. This includes short-lived collimated ejections from a precessing central source.
    
\section*{Data Availability}
This work makes use of public data from the Kinematic Catalogue of Planetary Nebula obtained at the San Pedro Mártir Observatory in Baja California, Mexico (\citealt{lopez2012san}).



\bibliographystyle{mnras}
\bibliography{References}


\bsp	
\label{lastpage}
\end{document}